\documentclass[12pt]{article}
\usepackage{amsfonts}
\usepackage{amssymb,letterspace}
\usepackage{graphics,psboxit,amsmath}
\usepackage{subfigure}
\usepackage{graphicx}

\def\hybrid{\topmargin -20pt    \oddsidemargin 0pt
        \headheight 0pt \headsep 0pt
        \textwidth 6.35in       
        \textheight 9.25in       
        \marginparwidth .875in
        \parskip 5pt plus 1pt   \jot = 1.5ex}

\hybrid

\def\baselinestretch{1.2}

\catcode`\@=11

\def\marginnote#1{}
\newcount\hour
\newcount\minute
\newtoks\amorpm
\hour=\time\divide\hour by60
\minute=\time{\multiply\hour by60 \global\advance\minute by-\hour}
\edef\standardtime{{\ifnum\hour<12 \global\amorpm={am}%
        \else\global\amorpm={pm}\advance\hour by-12 \fi
        \ifnum\hour=0 \hour=12 \fi
        \number\hour:\ifnum\minute<10 0\fi\number\minute\the\amorpm}}
\edef\militarytime{\number\hour:\ifnum\minute<10 0\fi\number\minute}

\def\draftlabel#1{{\@bsphack\if@filesw {\let\thepage\relax
   \xdef\@gtempa{\write\@auxout{\string
      \newlabel{#1}{{\@currentlabel}{\thepage}}}}}\@gtempa
   \if@nobreak \ifvmode\nobreak\fi\fi\fi\@esphack}
        \gdef\@eqnlabel{#1}}
\def\@eqnlabel{}
\def\@vacuum{}
\def\draftmarginnote#1{\marginpar{\raggedright\scriptsize\tt#1}}

\def\draft{\oddsidemargin -.5truein
        \def\@oddfoot{\sl preliminary draft \hfil
        \rm\thepage\hfil\sl\today\quad\militarytime}
        \let\@evenfoot\@oddfoot \overfullrule 3pt
        \let\label=\draftlabel
        \let\marginnote=\draftmarginnote
   \def\@eqnnum{(\theequation)\rlap{\kern\marginparsep\tt\@eqnlabel}%
\global\let\@eqnlabel\@vacuum}  }

\def\preprint{\twocolumn\sloppy\flushbottom\parindent 2em
        \leftmargini 2em\leftmarginv .5em\leftmarginvi .5em
        \oddsidemargin -.5in    \evensidemargin -.5in
        \columnsep .4in \footheight 0pt
        \textwidth 10.in        \topmargin  -.4in
        \headheight 12pt \topskip .4in
        \textheight 6.9in \footskip 0pt
        \def\@oddhead{\thepage\hfil\addtocounter{page}{1}\thepage}
        \let\@evenhead\@oddhead \def\@oddfoot{} \def\@evenfoot{} }

\def\numberbysection{\@addtoreset{equation}{section}
        \def\theequation{\thesection.\arabic{equation}}}

\def\underline#1{\relax\ifmmode\@@underline#1\else
        $\@@underline{\hbox{#1}}$\relax\fi}

\def\titlepage{\@restonecolfalse\if@twocolumn\@restonecoltrue\onecolumn
     \else \newpage \fi \thispagestyle{empty}\c@page\z@
        \def\thefootnote{\fnsymbol{footnote}} }

\def\endtitlepage{\if@restonecol\twocolumn \else \newpage \fi
        \def\thefootnote{\arabic{footnote}}
        \setcounter{footnote}{0}}  

\catcode`@=12
\relax

\def\figcap{\section*{Figure Captions\markboth
        {FIGURECAPTIONS}{FIGURECAPTIONS}}\list
        {Figure \arabic{enumi}:\hfill}{\settowidth\labelwidth{Figure
999:}
        \leftmargin\labelwidth
        \advance\leftmargin\labelsep\usecounter{enumi}}}
 \relax
\def\tablecap{\section*{Table Captions\markboth
        {TABLECAPTIONS}{TABLECAPTIONS}}\list
        {Table \arabic{enumi}:\hfill}{\settowidth\labelwidth{Table
999:}
        \leftmargin\labelwidth
        \advance\leftmargin\labelsep\usecounter{enumi}}}
 \relax
\def\reflist{\section*{References\markboth
        {REFLIST}{REFLIST}}\list
        {[\arabic{enumi}]\hfill}{\settowidth\labelwidth{[999]}
        \leftmargin\labelwidth
        \advance\leftmargin\labelsep\usecounter{enumi}}}
 \relax
\makeatletter
\newcounter{pubctr}
\def\publist{\@ifnextchar[{\@publist}{\@@publist}}
\def\@publist[#1]{\list
        {[\arabic{pubctr}]\hfill}{\settowidth\labelwidth{[999]}
        \leftmargin\labelwidth
        \advance\leftmargin\labelsep
        \@nmbrlisttrue\def\@listctr{pubctr}
        \setcounter{pubctr}{#1}\addtocounter{pubctr}{-1}}}
\def\@@publist{\list
        {[\arabic{pubctr}]\hfill}{\settowidth\labelwidth{[999]}
        \leftmargin\labelwidth
        \advance\leftmargin\labelsep
        \@nmbrlisttrue\def\@listctr{pubctr}}}
 \relax
\makeatother
\newskip\humongous \humongous=0pt plus 1000pt minus 1000pt

\newif\ifdtup

\relax

\def\be{\begin{equation}}
\def\ee{\end{equation}}
\def\ba{\begin{eqnarray}}
\def\ea{\end{eqnarray}}

\def\no{\noindent}

\def\IR{\relax{\rm I\kern-.18em R}}
\def\II{\relax{\rm 1\kern-.35em1}}

\renewcommand{\theequation}{\thesection.\arabic{equation}}
\csname @addtoreset\endcsname{equation}{section}

\def\IR{\relax{\rm I\kern-.18em R}}
\def\inv{^{\raise.15ex\hbox{${\scriptscriptstyle -}$}\kern-.05em 1}}

\begin{document}


\begin{titlepage}
\begin{center}

\hfill CERN-PH-TH/2005-032\\
\vskip -.1 cm
\hfill IFT-UAM/CSIC-05-14\\
\vskip -.1 cm
\hfill hep--th/0502188\\

\vskip .5in

{\LARGE Finite size effects in ferromagnetic
spin chains and quantum corrections 
to classical strings}
\vskip 0.4in

{\bf Rafael Hern\'andez$^1$}, {\bf Esperanza L\'opez}$^2$, \\
{\bf \'Africa Peri\'a\~nez}$^2$ and {\bf Germ\'an Sierra}$^2$
\vskip 0.2in

${}^1\!$
Theory Division, CERN \\
CH-1211 Geneva 23, Switzerland\\
{\footnotesize{\tt rafael.hernandez@cern.ch}}

\vskip .2in

${}^2\!$
Departamento de F\'{\i}sica Te\'orica C-XI
and Instituto de F\'{\i}sica Te\'orica  C-XVI\\
Universidad Aut\'onoma de Madrid,
Cantoblanco, 28049 Madrid, Spain\\
{\footnotesize{\tt esperanza.lopez, africa.periannez, german.sierra@uam.es}}

\end{center}

\vskip .4in

\centerline{\bf Abstract}
${\cal N}=4$ supersymmetric Yang-Mills operators carrying
large charges are dual to semiclassical strings in
$AdS_5 \times S^5$. The spectrum of anomalous dimensions of
very large operators has been calculated solving the
Bethe ansatz equations in the thermodynamic limit, and matched
to energies of string solitons. We have considered finite size corrections
to the Bethe equations, that should correspond to
quantum effects on the string side.

\no

\noindent

\vskip .4in
\noindent

\end{titlepage}
\vfill
\eject

\def\baselinestretch{1.2}


\baselineskip 20pt

\section{Introduction}

Very strong checks of the AdS/CFT correspondence beyond the supergravity regime 
have been obtained along the last years from the study of sectors with large 
quantum numbers, following ideas first presented in \cite{Polyakov:2001af}. 
Operators with a large R-symmetry charge $J$, of the form $\hbox{Tr } (Z^J \ldots)$, 
where $Z$ is one of the ${\cal N}=4$ complex scalars and the dots stand for 
insertions of few other fields, were mapped to small closed strings in $AdS_5 \times S^5$
whose center of mass moves with large angular momentum $J$ along a circle of 
$S^5$ \cite{BMN,GKP}. This analysis was afterwards extended to operators 
composed of the three ${\cal N}=4$ scalars, that were proposed to correspond to
semiclassical string solutions with three large angular momenta along $S^5$ \cite{FT}. 
The energy of these semiclassical strings admits an expansion in the effective 
coupling $\lambda/J^2$, with $\lambda$ the 't Hooft coupling of the gauge theory, 
suggesting the possibility of a precise comparison between string energies and 
anomalous dimensions of large ${\cal N}=4$ Yang-Mills operators. However operator 
mixing turned the computation of anomalous dimensions for large ${\cal N}\!=\!4$ 
operators into a formidable task, until the illuminating identification of the 
planar dilatation operator \cite{dilatation,Beisert} with the Hamiltonian of an 
integrable quantum spin chain \cite{MZ,psu}. The spectrum of anomalous dimensions 
for large operators became then computable using the powerful technique of the 
Bethe ansatz, and complete agreement with the energies of semiclassical string 
states was found at the first two leading orders in $\lambda$ \cite{spinning}-\cite{Ryang}. 
Moreover a perfect matching between the Bethe equations in the thermodynamic limit 
of very long spin chains and the integrability properties of the classical 
string on $AdS_5 \times S^5$ was shown up to order $\lambda^2$ in 
\cite{Kazakov}-\cite{n=4curve}. The correspondence was further reinforced by 
the direct comparison of the action describing the continuum limit of the spin 
chain in the coherent state basis with the dual string non-linear sigma model 
\cite{Martin}-\cite{fermions}. 

According to the AdS/CFT correspondence, finite size corrections to the
thermodynamic limit of very long spin chains provide quantum
corrections, beyond the classical limit, to strings carrying large
quantum numbers. The effect of finite size corrections on anomalous dimensions in the
$SU(2)$ sector, describing ${\cal N}\!=\!4$ operators composed of two complex scalars, was 
previously considered in \cite{LZ} and compared to one-loop corrections to classical 
string energies for circular strings \cite{FT} rotating in $S^5$ with two equal angular 
momenta \cite{semi,FPT}. In this case disagreement was found already at leading order in 
$\lambda$. The interpretation was however far from conclusive because the energy of the 
associated string contains an imaginary piece and the configuration is therefore unstable. 
A safer test was then proposed in \cite{SL2}, by studying the one-loop correction to a 
stable circular string rotating in both $AdS_5$ and $S^5$. The energy of this configuration 
corresponds to the anomalous dimension of an operator in the $SU(1,1)$ sector of 
${\cal N}=4$ Yang-Mills \cite{Beisert}. The correction to the thermodynamic limit of the Bethe equations in 
the $SU(1,1)$ sector was determined in \cite{SL2curve}. Again, as in the $SU(2)$ sector, 
a disagreement was found at order $\lambda$. Comparison of the one-loop quantum correction 
for a stable circular string with three angular momenta along $S^5$ to the finite size 
effect on the anomalous dimension for the dual operators was also recently considered 
in \cite{FK}. 

In this paper we will reconsider finite size corrections to the thermodynamic 
limit of the Bethe ansatz in the $SU(2)$ and $SU(1,1)$ sectors, and
argue that a crucial piece was missing in the previous analysis. The plan of the paper 
is the following. In section 2 we will introduce the Bethe ansatz
equations. In section 3 we will evaluate the first $\frac {1}{L}$ 
corrections to their thermodynamic limit. In section 4 we will compute the 
effect of finite size corrections on the energy of some particular solutions of the 
Bethe equations, which map to circular strings on $AdS_5 \times S^5$.
We conclude with some comments in section 5.


\section{The Bethe ansatz equations}

The isotropic spin 1/2 Heisenberg chain is one of the most studied 
integrable systems. At each site of the chain sits the fundamental
representation of $SU(2)$. We will be interested on studying this
spin chain in the ferromagnetic regime.
The ground state of the system, which spontaneously breaks the underlying 
$SU(2)$ symmetry, consists of all spins equally oriented at each site 
of the chain. Let us choose for definiteness 
$| \!\! \uparrow \uparrow \cdots \uparrow \rangle$.
The elementary excitations of the chain, called magnons, consist of one 
spin down over a sea of spins up carrying a definite momentum 
$\sum_{j=1}^L e^{ip j} |\uparrow \cdots \downarrow_j \cdots \uparrow \,
\rangle$, where $L$ is the number of sites of the chain.
The momentum $p$ is customarily parameterized as
\be 
e^{ip}={u+{i/ 2} \over u-{i/ 2}} \, ,
\ee
where $u$ is called the rapidity, which can be real
or complex. The spectrum of the chain
is given by collections of $M$ magnons, whose rapidities 
satisfy the Bethe ansatz equations, 
\be
\left( {u_j+{i/2} \over u_j-{i/2}} \right)^L =\;
\prod_{k \neq j}^M \, {u_j-u_k+i  \over u_j-u_k -i} \, .
\label{bethe}
\ee

${\cal N}=4$ Yang-Mills operators composed of two complex 
scalars, ${\rm Tr}(Z_1^{M}Z_2^{L-M})$, can be formally mapped to spin chain 
configurations by identifying $|\!\uparrow\rangle \rightarrow Z_1$
and $|\!\downarrow\rangle \rightarrow Z_2$. The order in which the fields appear 
inside the trace maps into the site arrangement of the spins along the chain. 
The cyclic invariance of the trace implies the restriction to cyclically invariant 
spin chain configurations, {\it i.e.} $e^{iP}=1$ where $P$ is the total momentum.  
It has been shown \cite{MZ} that the dilatation operator of ${\cal N}=4$ 
Yang-Mills on the holomorphic two scalar sector coincides with 
the Heisenberg hamiltonian.  

The Bethe ansatz equations \eqref{bethe} can be generalized to spin
chains with $SU(1,1)$  symmetry, in which case the LHS term
has to be replaced by $((u_j - i/2)/(u_j + i/2))^L$. 
The ${\cal N}=4$ counterpart of this chain corresponds to 
operators composed of a single complex scalar field and an
arbitrary number of derivatives, ${\rm Tr} (D^M Z^L)$. 
Both spin chains can be unified by representing
the lowering operator $S^-_i$ at the site $i$ 
as the product of two fermionic
destruction operators for $SU(2)$ or two bosonic destruction
operators for $SU(1,1)$. We shall use a parameter
$\epsilon$ to distinguish between the two options: 
$\epsilon = +1$ for  $SU(2)$ and $\epsilon = -1$ for $SU(1,1)$. 
These operator representations
are at the basis of another applications of the previous 
Bethe ansatz equations that we shall briefly comment below. 

We will be interested in chains with a large number of sites and
in solutions to the Bethe equations \eqref{bethe}
where the number of roots is comparable to the number of sites
of the chain, {\it i.e.} $L\gg 1$ and $M\sim L$. 
In the $SU(2)$ case, corresponding to $\epsilon = 1$, the roots 
form open arcs or ``strings'' 
in the rapidity $u$ complex plane, which are symmetric around the
real axis. The reason being  that if $u_j$ is a solution of \eqref{bethe}, 
so is $u^*_j=u_{k}$ for some $k$. 
In the $SU(1,1)$  case, $\epsilon = -1$,
the Bethe roots are real and condense in segments on the 
real axis \cite{RMP}. As in the limit $L\gg 1$ the Bethe roots scale as 
$u_j \sim L$, it is convenient to define a rescaled
variable $x \equiv {u \over L}$. Bethe strings with a 
macroscopic number of roots
$M \!\sim \! L$ locate along a well defined curve $\cal C$ in the $x$-plane, 
and describe macroscopic spin waves, dual to semiclassical strings in 
$AdS_5 \times S^5$. Along this curve a smooth function parameterizing 
the density of roots 
can be defined. Let us introduce an auxiliary real variable $n$, such that 
${\cal C}= \{ x(n), \, n \! \in [{1 \over L},{M \over L}] \}$ and 
$x\!\!\left({j \over L}\right)={u_j \over L}$.
We now define the function $\rho(x)$ as
\be
\rho^{-1}={dx \over dn} \, .
\label{density}
\ee
In general this function is complex-valued, unless $\cal C$ lies along the 
real axis. The density of Bethe roots along $\cal C$ is then given by 
$|\rho(x)|$. Notice that $|\rho(x_j)|\!\sim\!1$ implies separations
between neighbouring roots $|u_j\!-\!u_k|\!\sim\!1$. The function 
$\rho(x)$, as given in \eqref{density}, should be interpreted as a 
smooth interpolating function. It is not directly equivalent to the definition 
commonly used in the literature, see for example \cite{Kazakov}, where
$\rho$ is given by a sum over delta functions at the positions of the
Bethe roots. However, both definitions coincide in the thermodynamic limit 
of infinitely long spin chains. 

From now on we will concentrate in solutions to the Bethe equations 
consisting of several macroscopic strings. In the limit $L\gg 1$, the Bethe 
equations are best analyzed in the logarithmic form
\be
-i L \, \log \, {u_j+{i/2} \over u_j-{i/2}} \; + \; 2 \pi n_j \; =\;-i 
\sum_{k \neq j} \, \log \, {u_j-u_k+i  \over u_j-u_k -i}  \, .
\label{bethelog}
\ee
The integers $n_j$ originate from the multivaluedness of the logarithm, and 
label different Bethe strings. The type of solutions we are interested in corresponds 
to having the roots distributed along those sets for which the integers $n_j$
coincide, each set containing a number of roots of order $L$. In the thermodynamic 
limit where $L \rightarrow \infty$ and $M \rightarrow \infty$, 
with the ratio $M/L$ kept fixed, the equations \eqref{bethelog} become integral 
equations for the density of roots along each curve ${\cal C}_s$
\be
{\epsilon \over x} 
\,+\, 2\pi n_s \;= \;2 \, - \hspace{-5mm}\int_{{\cal C}_s} dx' {\rho_s(x') \over x -x'}
\;+ \;2 \sum_{r\neq s} \int_{{\cal C}_r} dx' {\rho_r(x') \over x -x'} \, ,
\hspace{.5cm} x \in {\cal C}_s .
\label{betheint}
\ee
  
Relation \eqref{betheint} is obtained assuming that, since $u_j \sim L$,
in the thermodynamic limit the logarithms can be replaced by the leading
term in their Taylor expansion on inverse powers of $u$. Since there
is a macroscopic number of roots, summations are then substituted by 
integrals. Let us analyze the previous assumption in some detail.
The curves ${\cal C}_s$ on the $x$-plane along which the solutions of equation  
\eqref{betheint} distribute have in general finite size.
Each of them accommodates a large 
number of roots, of order $L$, and the density 
tends to zero at the endpoints. Notice also that 
the density of roots scales as $\rho_s \sim n_s$.
Both these facts indicate that inside the curve we can find densities 
close to or even larger than one. Thus there will be regions where the distance 
between neighbouring roots is $|u_j-u_k| \lesssim \!1$. Roots for which this happens 
do not allow for a power series expansion of the logarithms in equation \eqref{bethelog} 
or at least do not justify to keep only the leading term. Since the number of roots 
close to a given one where this situation arises is expected to be of order one, 
they should not contribute to the leading behaviour in the thermodynamic limit. However 
these considerations become relevant when analyzing the finite size corrections 
to \eqref{betheint}.

Before we analyze these corrections we shall make some general comments 
concerning the equations \eqref{betheint}. First of all these equations 
admit an electrostatic analogue as the equilibrium conditions for a set 
of point like charges interacting among themselves ($1/(x - x')$ terms)
and with a charge located at the origin ($1/x$ term). The $2 \pi n_s$ 
term can then be interpreted as a constant electric field, which depends 
on the given curve ${\cal C}_s$ \cite{R-limit}-\cite{RMP}. Equations like 
\eqref{betheint} have also appeared in a different context specially in the 
study of exactly solvable BCS models of superconductivity \cite{RS}, the 
Gaudin magnets \cite{G-book}, bosonic BCS models \cite{R-boson,Jorge-boson} 
and the Russian doll BCS model \cite{RD1,RD2}. In particular the Bethe 
equation of the BCS model for a single energy level with high degeneracy
coincides precisely with \eqref{betheint}. This is in fact the simplest 
BCS model and can be seen as the limit of any BCS model where the BCS 
coupling constant is much larger than the remaining energy scales of the 
model, as for example the Debye energy. The possibility of having different
arcs for the solutions of the Bethe equations of the BCS model was considered
by Gaudin in  \cite{G-book} but they have not found so far any practical
application in superconductivity or condensed matter. The case of only
one arc coincides in the thermodynamic limit with the standard mean field 
solution of the BCS model to leading order in the number of particles.  


\section{Finite size corrections}

In this section we will evaluate the leading $1 \over L$ corrections to
the Bethe equations. If the curves ${\cal C}_s$ where the roots condense 
do not contain the point $x=0$, we can always approximate the logarithm in the LHS of 
equation \eqref{bethelog} by the first term in its series expansion. Thus the 
LHS of \eqref{betheint} is not modified in the limit of large but
not infinite $L$. We will focus then in the RHS of equation \eqref{bethelog}.
When two roots $x_j$, $x_k$ lay on different curves, we can assume
that their separation is of order one and again the associated
logarithm on the RHS of equation \eqref{bethelog} can be well approximated by
the first term in its series expansion. The only problematic
situation arises when we consider roots belonging to the same 
curve. Hence we can restrict to the simplest situation on which
all the roots condense to form a single curve $\cal C$ on the 
$x$-plane. The quantity whose finite size corrections we have to
determine is
\be
-i \sum_{k \neq j} \, \log \, {x_j-x_k+i/L  \over x_j-x_k -i/L}  \; .
\label{sum}
\ee
  
Given a root $x_j \! \in \! {\cal C}$, let us choose two integer 
$N_1$, $N_2$ with the property that
\ba
& \!\!\!\!\!\! |x_j - x_{j \pm l}| > {1 \over L}  & 
\;\; {\rm if}  \hspace{4mm} l \geq N_1 \;
\label{cond}, \\
& \! \! |x_j - x_{j \pm l}| > \delta
& \;\; {\rm if} \hspace{4mm} l \geq N_2 \; , \nonumber
\ea
with $\delta \!<\!<\!1$, but of order $L^0$.
Since $\rho(x)$ is a smooth function on $\cal C$, we can then further 
ask that $N_1$ is of order $L^0$ and $N_2$ of order $L$.
We divide then the summation in \eqref{sum} into three
separate regions

\vspace{4mm}
\begin{figure}[h!]
\hspace{4mm}
\includegraphics[height= 3 cm,angle= 0]{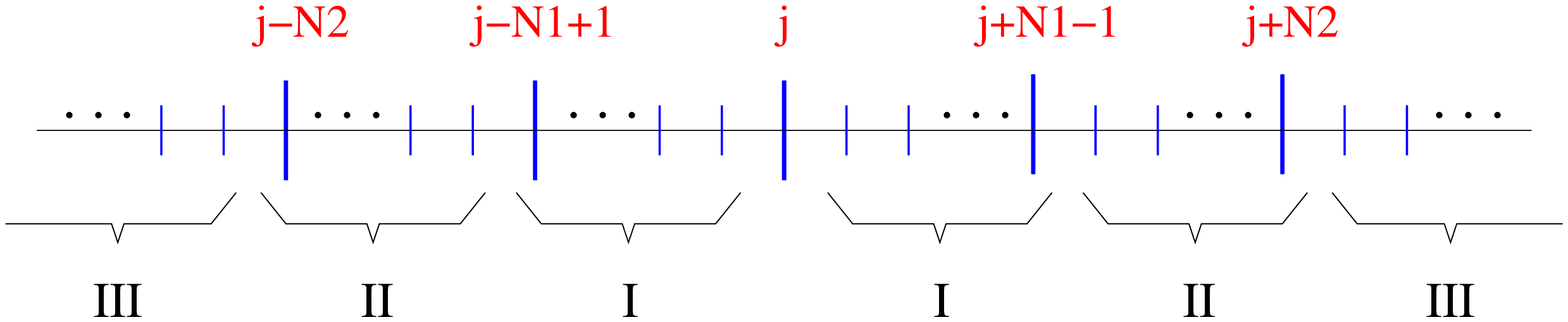}
\end{figure}

The separation between roots is infinitesimal in region I, and finite but
very small in region II. In both cases it can be evaluated explicitly as follows,
\be
x_{j+l}-x_j= \int_{n_j}^{n_{j+l}} {dn \over \rho} \approx
\int_{n_j}^{n_{j+l}} {dn \over \rho_j} \left[ 1- {\rho'_j \over
\rho_j^2}(n-n_j) \right]= {l \over L \rho_j}- {\rho'_j \over 2 \rho_j}
\left({l \over L \rho_j}\right)^2 \, ,
\label{dif}
\ee
where $n_j\!=\!{j \over L}$ and $n_{j+l}\!=\!n_j\!+\!{l \over L}$, 
and to simplify notation we have defined $\rho_j\!=\!\rho(x_j)$ and 
$\rho'_j\!=\!{d\rho \over dx}(x_j)$. This relation will be the main 
tool to determine the leading finite size corrections to the Bethe 
equations.

Let us analyze first the contribution from region I to 
\eqref{sum}. It can be most conveniently written as
\be
-i \sum_{l=1}^{N_1-1} \left[\, \log \, {x_j-x_{j-l}+i/L  
\over x_{j+l}-x_j +i/L} \;
- \, \log \,{x_j-x_{j-l}-i/L  \over x_{j+l}-x_j -i/L} \, \right] \, .
\label{sumI}
\ee
Region I has been introduced to take into account possible situations
where $L|x_j\!-\!x_k|<\!1$. This can happen typically on the central 
part of the condensation curve $\cal C$, where the density attains its
maximum. On the other hand, depending on the choice of $N_1$, this 
region will contain a number of roots 
with $L|x_j\!-\!x_k|\sim \!1$. Therefore it is not possible 
to approximate the logarithms in \eqref{sumI} by a power
expansion. However that expression can be easily analyzed with
the help of \eqref{dif}. A non-vanishing result is obtained due to the
sub-leading contribution to the separation between roots given by the 
last term in \eqref{dif}. Substituting in \eqref{sumI} 
and neglecting terms ${\cal O}(L^{-2})$, we get the following result 
\be
- {2 \, \rho'_j 
\over L \rho_j} \; \sum_{l=1}^{N_1-1} {l^2 \over l^2 + \rho_j^2} \; .
\label{sumlog}
\ee

We turn now to region II. Since in this region $L|x_j\!-\!x_k|>\!1$, 
we can represent the logarithm by the series expansion 
\be
-i \log \, {x_j-x_k +i/L \over x_j-x_k -i/L}= 2 \sum_{n=0}^\infty
{(-1)^n \over (2n+1) [L(x_j-x_k)]^{2n+1}} \; .
\label{exp}
\ee
The term with $n=0$ provides the main contribution to \eqref{sumI}.
The terms with $n\!>\!0$ are naively suppressed
by powers of $L^{-2}$ with respect to the term with $n\!=\!0$. 
However, since region II contains roots with $L|x_j\!-\!x_k|\sim\!1$, 
it is not justified to neglect them. Thus, let us study the 
contribution from each separate term $n\!>\!0$ in the expansion of the sum
\eqref{sum}, which we write as
\be
2 \sum_{l=N_1}^{N_2} {(-1)^n  \over(2n+1)}
\left[{1 \over [L(x_j-x_{j-l})]^{2n+1}} - 
{1 \over [L(x_{j+l}-x_j)]^{2n+1}}\right] \; .
\ee
This expression can be evaluated using once more \eqref{dif}, giving the
simple result
\be
- {2 \rho'_j  \over L} \;(-1)^n  \rho_j^{2n-1}  \sum_{l=N_1}^{N_2}
{1 \over l^{2n}} \; .
\label{taylor}
\ee
Since $N_2$ is of order $L$, one should check that the
approximate formula \eqref{dif} captures all the leading finite size
corrections. Higher orders in the expansion \eqref{dif} would
lead to an expression analogous to \eqref{taylor}, but
multiplied by additional powers of $\big({l \over L}\big)^2$. 
The most dangerous situation arises when $n\!=\!1$. Given that for 
this case the sum in \eqref{taylor} goes like $l^{-2}$, the first
additional correction is
proportional to ${1 \over L^3} \sum 1$. It is thus suppressed as $L^{-2}$. 
Notice that a different conclusion is obtained for the term $n\!=\!0$ in the
expansion of the logarithm, for which
the next contribution in the expansion \eqref{dif} would produce a piece
${1 \over L^3} \sum l^2 \sim L^0$. Therefore the leading term must be treated
independently.

As the next step, we extend the upper limit of the summation in
\eqref{taylor} from $N_2$ to 
infinity. This does not affect the leading finite size corrections. 
Indeed, the stronger effect of 
this substitution happens again for $n\!=\!1$ and
it is of order $L^{-{2}}$. 
Using that ${\rho_j \over l}\!<\!1$ in region II, the sum over $n\!>\!0$
of the terms \eqref{taylor} can be recognized as the Taylor 
expansion that reconstructs the function
\be
{2 \rho_j \rho_j' \over  L} \; \sum_{l=N_1}^{\infty} {1 \over l^2 + \rho_j^2} 
\; .
\ee
It is interesting to compare the contribution from region I,
given by expression \eqref{sumlog}, with what would have been obtained
from evaluating in that region the term $n\!=\!0$ in \eqref{exp}.
Now relation \eqref{dif} can be used
without problems, and we observe that the difference between both
quantities is just 
\be
{2 \rho_j \rho'_j \over L} \;\, \sum_{l=1}^{N_1-1} {1 \over l^2 + \rho_j^2} 
\; .
\ee

Clearly only the term $n\!=\!0$ contributes to the leading finite size 
effects in region III.
Collecting all contributions we obtain the final result
\be
-i \, \sum_{k \neq j} \log\, {x_j-x_k+i/L \over x_j -x_k -i/L} \; = \;
{1 \over L}\, \sum_{k \neq j} {2 \over  x_j -x_k}\; + \;
{2 \rho_j \rho_j' \over L} \, 
\sum_{l=1}^{\infty} {1 \over l^2 + \rho_j^2} \; ,
\label{sumF}
\ee
up to terms of order $L^{-2}$. As required for consistency, this 
relation is independent of $N_1$ and $N_2$. It shows that approximating 
the logarithm by the first term in its Taylor expansion is valid in the 
thermodynamic limit, but it is not enough to capture the leading finite 
size effects. It is remarkable that the last term in \eqref{sumF} also 
represents of a summation over roots, where each summand gives
a contribution from the roots $x_{j\pm l}$ around the chosen one $x_j$. 
Using that 
\be
\pi \rho \coth \pi \rho = \sum_{l=-\infty}^\infty 
{\rho^2 \over l^2 + \rho^2} \; , 
\ee
the previous relation can be written in the more compact form
\be
-i \, \sum_{k \neq j} \log\, {x_j-x_k+i/L \over x_j -x_k -i/L} \;=\;
{1 \over L} \sum_{k \neq j} {2 \over x_j -x_k} + {\rho'_j \over \rho_j L} 
\left(\pi \rho_j \coth \pi \rho_j -1 \right) \; . 
\label{fsize}
\ee
Since we are only interested in the leading finite size corrections,
the function $\rho$ should be understood in the strict
thermodynamic limit. In figure 1 we show a numerical check of \eqref{fsize}.
An alternative derivation of this formula consists in 
making an infinitesimal variation of the positions $x_j$ 
along the arc ${\cal C}$. Using \eqref{dif} one 
can then relate $\delta x_k - \delta x_l$  with a variation of the density
function $\delta \rho_j$ and its derivative $\delta \rho'_j$. 
It is easy to check that both sides coincide up to higher powers
in $1/L$. 

\begin{figure}[t!]
\hspace{-7mm}
\subfigure{\includegraphics[height= 6 cm,angle= 0]{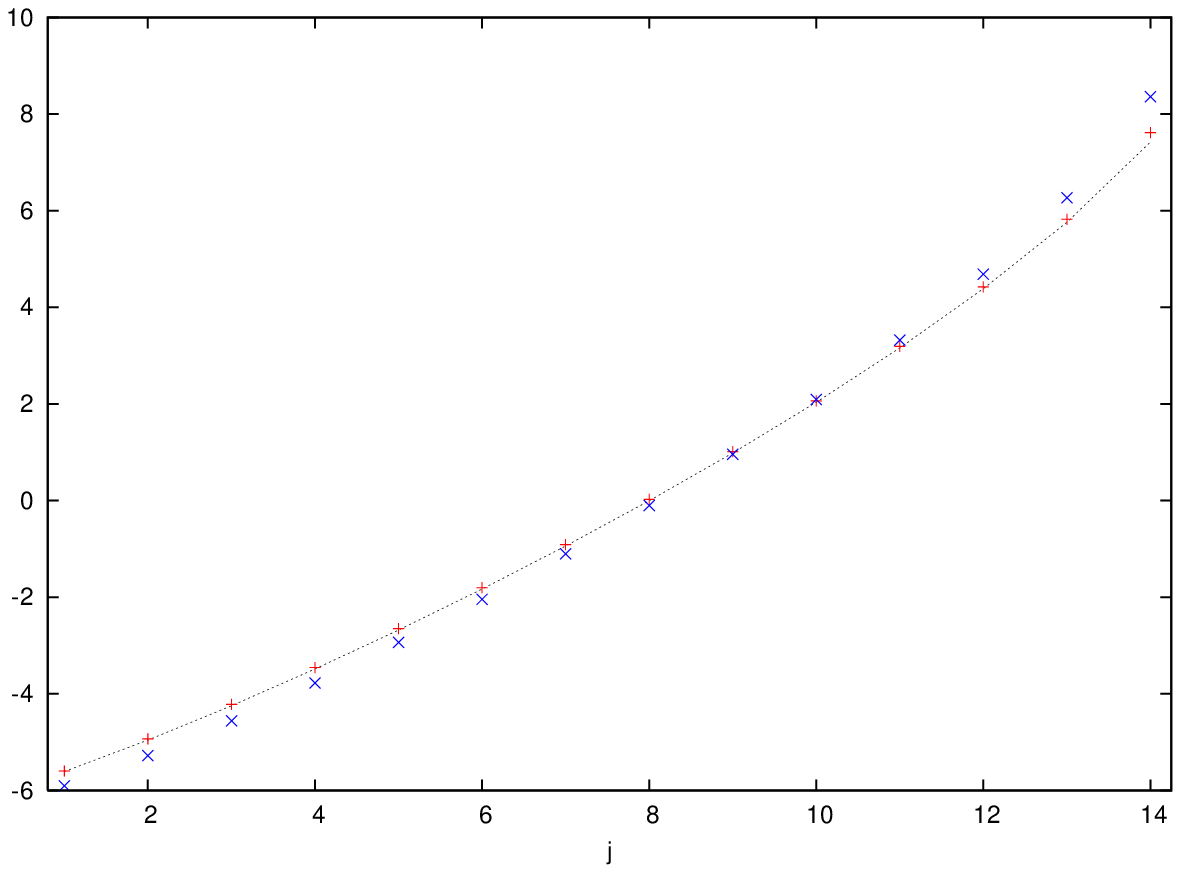}}
\subfigure{\includegraphics[height= 6 cm,angle= 0]{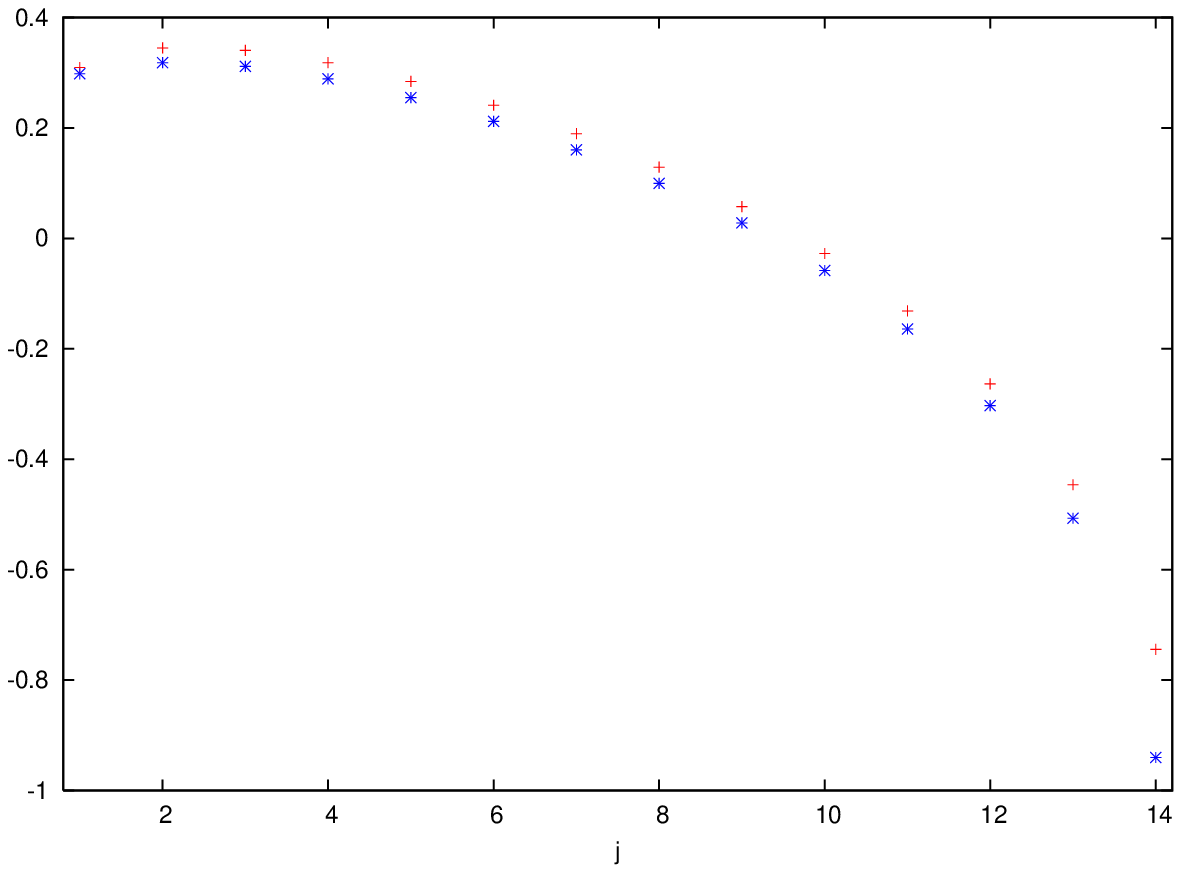}}
\caption{
Check of \eqref{fsize} for a density $\rho\! =\! 
3x \sqrt{1\!-x^2}$ for which $x_j\!\! =\!\!
\sqrt{1-(1\!-\!j/L)^{2/3}}$ with $j=1,..,L\!-\!1$ and
$L=15$. $(a)$ The dotted line is the LHS of
\eqref{fsize}, while the $+$ are the RHS; the $\times$ denote the 
first term in the RHS. $(b)$ The $\ast$ represent the
difference between the LHS and the first term in the
RHS of \eqref{fsize}, and the $+$ are the $1/L$ correction.}
\label{fig2}
\end{figure}

In the general situation on which the Bethe roots condense on several
macroscopic curves, \eqref{fsize} will apply to the summation over roots 
belonging to each separate curve. When the roots $x_j$, $x_k$ lay on 
different curves, the last term on \eqref{fsize} is absent. 
Let us denote by $x_{s,j}$ the roots belonging to the $s$-th curve.
The integral Bethe equations \eqref{betheint}, 
valid in the limit of an infinite chain, gets modified as follows by the first
finite size corrections
\ba
{\epsilon \over x_{s,j}} \;+\;2 \pi n_s & = &\label{bethe1overL} 
{1 \over L} \, \sum_{k\neq j} {2 \over x_{s,j}-x_{s,k}} \; +
\label{bL} \\
&+& {1 \over L} \, \sum_{r\neq s}\sum_{k} {2 \over x_{s,j}-x_{r,k}} 
\; +\; {\rho'_{s,j} \over \rho_{s,j} L} \nonumber 
\left(\pi \rho_{s,j} \coth \pi \rho_{s,j} -1 \right) \; .
\ea 
These equations can be straightforwardly generalized to
spin chains based on general Lie groups. 

We would like to end this section with a comment on the 
validity of \eqref{fsize} and \eqref{bL}. In deriving these equations,
we have assumed that the number of roots around the chosen one 
$x_{s,j}$ is of order $L$. This stops being true for 
$x_{s,j}$ sufficiently close to the end points of the associated 
condensation curve. We expect that this problem does not
affect the $1/L$ corrections to the conserved
charges of the macroscopic Bethe solutions. However it will
be important to estimate the order of the next to leading
finite size effects. Let us look again to figure 1.
In our example, the roots $x_j$ lay on the interval $[0,1]$. 
For $x$ close to $1$ the density is $\; \sim \!\sqrt{1-x}$,
reproducing the typical behaviour of the densities which solve the
Bethe equations. In figure 1b we observe that the $1/L$ correction 
that we have obtained is a worse approximation for $x$ close to $1$. 
In spite of that, this problem is practically confined to the last root.

It has been argued that quantum corrections to 
the conserved charges of spinning strings on $AdS_5 \times S^5$ 
are suppressed by integer powers of $1/L$ \cite{semi,Tseytlin:2003ii}.
According to the AdS/CFT correspondence, quantum corrections to 
semiclassical strings map to finite size 
effects for the spin chain describing the spectrum of anomalous dimensions
of ${\cal N}=4$ Yang-Mills. Hence it is interesting that, in addition to 
evaluating the leading finite size corrections for 
roots not very close to the end points, we were able to show that for
them the next to leading effects are of order $1/L^2$.


\section{Finite size corrections to the energy}

We will now use the results of the previous section to calculate the
leading finite size corrections to the energy of certain solutions to the 
Bethe equations for $SU(2)$ and $SU(1,1)$ spin chains.
The energy is in both cases given by
\be
E= {\lambda \over 8 \pi^2} \sum_{j=1}^M {1 \over u_j^2+ 1/4} \approx
{\lambda \over 8 \pi^2 L^2} \sum_{j=1}^M {1 \over x_j^2} \; .
\label{e}
\ee
We will concentrate again in solutions of the Bethe equation where
the roots condense on a single curve $\cal C$. The AdS/CFT
correspondence relates these solutions to the simplest and most
studied example of semiclassical strings spinning on $AdS_5 \times S^5$, {\it i.e.} 
circular strings \cite{FT,semi,FPT,SL2}. The $SU(2)$ and $SU(1,1)$
cases correspond respectively to strings rotating with two angular momenta 
on $S^5$, or both on $S^5$ and $AdS_5$. Comparing the energies on both sides 
of the correspondence it will be possible to determine
if the perfect agreement found in the thermodynamic/classical limit 
\cite{Kazakov}-\cite{n=4curve}, extends to the finite size/quantum level.  
  
The Bethe equations \eqref{bethe1overL} reduce in the case of a single
curve to
\be
{\epsilon \over x_j} \,+\,2\pi n \;=\; {1 \over L} \,\sum_{k\neq j} 
{2 \over x_j-x_k}\;+\;  {\rho'_{j} \over \rho_{j} L} 
\left(\pi \rho_{j} \coth \pi \rho_{j} -1 \right) \; ,
\label{rat}
\ee
where $\epsilon=1,-1$ for the $SU(2)$ and $SU(1,1)$ cases, respectively.
Defining $\alpha\!=\!{M \over L}$, the function $\rho(x)$ is given by
\cite{Kazakov,SL2curve,bosonBAE}
\be
\rho(x)\,=\,{\sqrt{8 \pi n \alpha x -(2 \pi n x + \epsilon)^2} \over 2 \pi x}
\; . \label{denrat}
\ee

Summing \eqref{rat} over $j$ we obtain
\be
\sum_{j=1}^M {\epsilon \over x_j}\,=\, -2 \pi n M 
\;+  \int_{\Gamma} d \rho \left(\pi \rho \coth \pi 
\rho -1 \right) \; ,
\label{poles}
\ee
where $\Gamma$ is the contour on the $\rho$-plane defined by
the function $\rho(x)$, with $x\!\in \!{\cal C}$. Since on the
end points of $\cal C$ the density is zero, the contour $\Gamma$
is closed and always contains the origin. For $SU(1,1)$ the curve $\cal C$ lies
on the real axis and $\rho(x)$ is a real function. $\Gamma$ should
be understood then as a closed contour surrounding the interval
between the origin and the maximal value of the density, 
$\rho_M$ (see figure 2a). Hence the integral on \eqref{poles} clearly vanishes. 
For $SU(2)$ the contour $\Gamma$ is symmetric under reflexion along the
imaginary axes, to which it cuts at zero and the maximal value of the density, 
$|\rho_M|$ (see figure 2b). Depending on the values of $n$ and $\alpha$ this 
contour can encircle some of the poles of the integrand in \eqref{poles}, which lie at 
$\rho=i \mathbb{Z}^\ast$, implying that the integral does not vanish.
We will not enter here the study of this more involved situation and its 
analysis in the context of the AdS/CFT correspondence. In this section
we will consider only $SU(2)$ solutions for which $|\rho_M|\!<\!1$
and thus the integral on \eqref{poles}, as for the $SU(1,1)$ case, vanishes.

\begin{figure}[t!]
\begin{center}
\includegraphics[height= 5 cm,angle= 0]{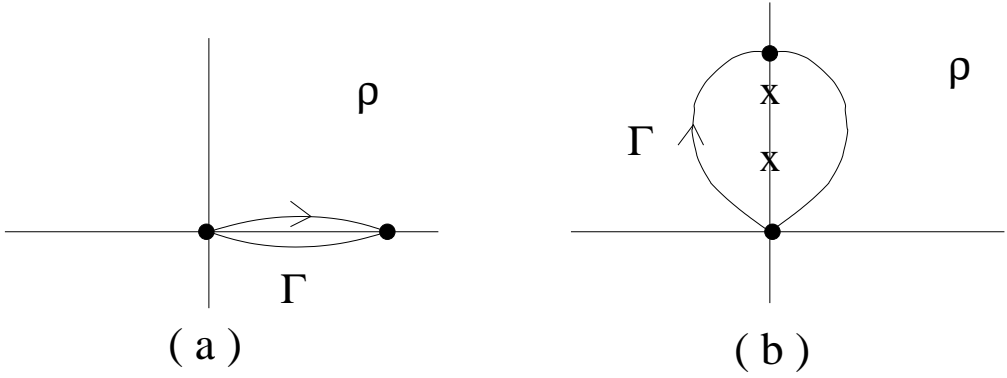}
\end{center}
\caption{Contour $\Gamma$ for a) $SU(1,1)$ and b) $SU(2)$ cases.  
The dots, $\bullet$, are the points $\rho=0, \rho_{M}$. The crosses, 
$\times$, in figure b)  represent the poles of $\pi \rho \coth \pi \rho$.   
}
\label{fig3}
\end{figure}
 
Multiplying \eqref{rat} by $\epsilon /x_j$ and then summing we get
\be
\sum_{j=1}^M {1 \over x_j^2} \,=\, 4 \pi^2 n^2 M 
\;+\; {\epsilon \over L}\, 
\sum_{k\neq j} {2 \over x_j(x_j-x_k)} \;+ \; \epsilon
\int_{\Gamma} {d\rho \over x(\rho)} \big( \pi \rho \coth \pi \rho-1
\big) \; .
\label{esum}
\ee
The function $x(\rho)$ is obtained inverting \eqref{denrat}
\be
x\,=\,{(2\alpha -\epsilon)\,n - \sqrt{4 \,
n^2\alpha(\alpha\!-\!\epsilon) -\rho^2} \over 2\pi (\rho^2+n^2)} \; .
\label{puz}
\ee
This is a double valued function with branch points at $\pm \rho_\ast$,
with $\rho_\ast =2 n \sqrt{\alpha(\alpha - \epsilon)}$. Since the
condensation curve lies on the real axis for the $SU(1,1)$ case,
it is clear that the maximal value of the density is attained at the branch point,
$\rho_M=\rho_\ast$. The situation is rather different for $SU(2)$. The curve $\cal C$ forms an
arc on the $x$-plane symmetric under reflexion along the real axis,
whose precise location is determined by the condition that $\rho(x)dx$
should be real. This translates into a transcendental equation that can be
analyzed numerically \cite{largeN}. The maximal value of the density is 
attained at the intersection point between $\cal C$ and the real axis, and it
turns out to verify $|\rho_M|>|\rho_\ast|$. The difference between $\rho_M$ 
and $\rho_\ast$ is maximal for the case of half-filling
while for small $\alpha$, $\rho_M \rightarrow \rho_\ast$. 

The spin chain dynamics can be alternatively described in the language
of coherent states. The coherent states which 
correspond to the $SU(2)$ single curve solutions have been shown to be
unstable when $|\rho_\ast|>1$ \cite{su3}. This condition does have a 
translation in terms of the Bethe solutions: some poles of the integrand 
in \eqref{poles} lie between the branch points of the function $x(\rho)$.
The fact that there are two different distinguished values of the density, 
$\rho_M$ and $\rho_\ast$, deserves a better understanding. It is important 
to stress that all $SU(2)$ solutions that can be mapped to ${\cal N}=4$ Yang-Mills 
operators, or equivalently, to semiclassical strings on $AdS_5 \times S^5$ 
are unstable and thus have $|\rho_M|>1$. Hence they are excluded from the 
analysis below, which is restricted for the $SU(2)$ case to $|\rho_M|<1$.

The second term on the RHS of \ref{esum} is easily evaluated to be
\be
\sum_{k\neq j} {2 \over x_j(x_j-x_k)}= -(2 \pi n M)^2+ 
\sum_{j=1}^M {1 \over x_j^2} \, .
\ee
Using \eqref{e} we obtain for the leading contribution to the energy 
\be
E_0= {\lambda \over 2L} n^2 \alpha(1-\epsilon \alpha) \; ,
\ee
which reproduces the results of \cite{Kazakov,SL2curve}. 
It is equivalent but more transparent to write the integral in
\eqref{esum} in terms of a contour ${\bar \Gamma}$ surrounding the branch
cut of the function $x(\rho)$, lying between 
$\, \pm 2 n \sqrt{\alpha(\alpha - \epsilon)}$. Then 
the leading finite size correction to the energy is
\be
E_1= {\epsilon \lambda \over 2L^2} \left[ n^2 \alpha(1-\epsilon \alpha) 
 \;+ \;{1 \over 8 \pi^2}
\int_{\bar \Gamma} {d \rho \over x(\rho)} \big( \pi \rho \coth \pi \rho-1
\big) \, \right]\; . 
\label{e1}
\ee
The first term comes from the $n\!=\!0$ piece in the approximation
of the logarithm by a series expansion. It gives the contribution 
to the finite size effects calculated in \cite{LZ,SL2curve}. Contrary 
to the $SU(2)$ circular strings, the $SU(1,1)$ ones are stable. In \cite{SL2} 
the leading quantum correction to their classical energy was calculated, 
finding that the first term in \eqref{e1} only takes care of the zero mode 
fluctuations. A. Tseytlin has informed us that the last term in \eqref{e1}
restores the perfect agreement with the string results. This represents one 
of the strongest tests of the AdS/CFT correspondence obtained up to now.
  
It is interesting to compare expression \eqref{e1} with the finite
size corrections to the ground state energy $E(L)$ of a critical 
Hamiltonian $H$ defined in a strip of width $L$, which is given 
by $E_0(L) = - \pi c/(6 L)$ where $c$ is the Virasoro central charge \cite{cardy}.
A well known example is the antiferromagnetic Heisenberg Hamiltonian, 
where $c=1$, which is at the heart of the bosonization approach to 
this spin chain \cite{boson}. In a relativistic $1+1$ model, where 
$E \sim k$, one can exchange time and space. Therefore the previous 
formula can also be seen as the free energy per unit length $F/(k_B T)$, 
which yields a low temperature specific heat $C \sim \pi c k_B^2 T/3$.
In the case of equation \eqref{e1} the $1/L$ correction does not
have such a CFT interpretation, since for the dual string it simply corresponds 
to the classical energy. However, the quantum corrections of the string,
given by the zero point energy, appear in the next order correction, 
${\cal O}(L^{-2})$, to the energy and in that sense are the analogue of 
the $c$ term. This is related to the fact that a ferromagnetic system has a dispersion
relation $E  \sim k^2$, so that time, which is an inverse temperature, 
should be exchanged by (space)$^2$. Hence we expect a correspondence between the finite 
size effects for the ferromagnetic Hamiltonian and the finite temperature effects of the 
system in infinite size, where all the low energy excitations contribute.   


\section{Conclusions}

In this paper we have considered finite size corrections to the thermodynamic limit of the   
Bethe ansatz for the $SU(2)$ and $SU(1,1)$ ferromagnetic spin chains. Anomalous dimensions 
of large ${\cal N}=4$ Yang-Mills operators can be computed by solving the Bethe ansatz 
for the corresponding spin chain, and compared with the energies of strings with two 
angular momenta rotating entirely along $S^5$ in the $SU(2)$ case, or both along 
$AdS_5$ and $S^5$ in the $SU(1,1)$ sector. Therefore our computation of the finite 
length effects provides the leading order 
correction, in $1/L$, to the anomalous dimensions of the operators. The finite size 
effects correspond, on the dual string theory side, to quantum corrections to classical 
string solutions. Hence the first $1/L$ correction that we have found
represents, following the AdS/CFT correspondence, the one-loop 
correction to the string energy.
  
The finite size Bethe ansatz equations that we have obtained, including the $1/L$ correction, may allow for a 
comparison with the string dynamics along the lines of \cite{Kazakov,SL2curve,Beisert:2004ag}. 
This would imply not only the matching with the energy, but also with the complete tower 
of conserved charges. The extension of the perfect agreement between the integrable
structure associated to long Yang-Mills operators and that of
classical strings on $AdS_5 \times S^5$ to the first quantum corrections would be
a most remarkable test of the AdS/CFT correspondence.


\newpage

\centerline {\bf Acknowledgments}

It is a pleasure to thank A. Tseytlin for numerous discussions and informing us 
of their work prior to publication \cite{BTZ}. We are also grateful to C. G\'omez 
and K. Zarembo for comments and correspondence. E.~L. was supported by a 
Ram\' on y Cajal contract of the MCYT and in part by the Spanish DGI under 
contract FPA2003-04597. A.~P. was supported by a Spanish fellowship FPI. Finally 
G.~S. has been supported by Spanish DGES under contract BFM2003-05316-C02-01. We 
also thank the EC Commission for financial support via the FP5 Grant
HPRN-CT-2002-00325. \vspace{-10 pt}



\begin{thebibliography}{99}

\renewcommand{\baselinestretch}{1}
\normalsize

\bibitem{Polyakov:2001af} A.~M.~Polyakov,
{\it Gauge fields and space-time},
Int.\ J.\ Mod.\ Phys.\ A {\bf 17S1} (2002) 119,
{\tt hep-th/0110196}.

\bibitem{BMN} D.~Berenstein, J.~M.~Maldacena and H.~Nastase,  
{\it Strings in flat space and pp waves from N = 4 super Yang Mills},
JHEP {\bf 0204} (2002) 013, {\tt hep-th/0202021}.

\bibitem{GKP} S.~S.~Gubser, I.~R.~Klebanov and A.~M.~Polyakov,  
{\it A semi-classical limit of the gauge/string correspondence},
Nucl.\ Phys.\ B {\bf 636} (2002) 99, {\tt hep-th/0204051}.

\bibitem{FT} S.~Frolov and A.~A.~Tseytlin,
{\it Multi-spin string solutions in $AdS_5 \times S^5$},
Nucl.\ Phys.\ B {\bf 668} (2003) 77, {\tt hep-th/0304255}.
  
S.~Frolov and A.~A.~Tseytlin,
{\it Rotating string solutions: AdS/CFT duality in non-supersymmetric 
sectors},
Phys.\ Lett.\ B {\bf 570} (2003) 96, {\tt hep-th/0306143}.

G.~Arutyunov, S.~Frolov, J.~Russo and A.~A.~Tseytlin,
{\it Spinning strings in $AdS_5 \times S^5$: New integrable system relations},
Nucl.\ Phys.\ B {\bf 671} (2003) 3, {\tt hep-th/0307191}.

G.~Arutyunov, J.~Russo and A.~A.~Tseytlin,
{\it Spinning strings in $AdS_5 \times S^5$ and integrable systems},
Nucl.\ Phys.\ B {\bf 671} (2003) 3, {\tt hep-th/0311004}.

\bibitem{dilatation} N.~Beisert, C.~Kristjansen and M.~Staudacher,
{\it The dilatation operator of ${\cal N} = 4$ super Yang-Mills theory},
Nucl.\ Phys.\ B {\bf 664} (2003) 131, {\tt hep-th/0303060}.

\bibitem{Beisert} N.~Beisert,
{\it The complete one-loop dilatation operator of ${\cal N} = 4$ super Yang-Mills theory},
Nucl.\ Phys.\ B {\bf 676} (2004) 3, {\tt hep-th/0307015}.

\bibitem{MZ} J.~A.~Minahan and K.~Zarembo,
{\it The Bethe-ansatz for ${\cal N} = 4$ super Yang-Mills},
JHEP {\bf 0303} (2003) 013, {\tt hep-th/0212208}.

\bibitem{psu} N.~Beisert and M.~Staudacher,
{\it The ${\cal N} = 4$ SYM integrable super spin chain},
Nucl.\ Phys.\ B {\bf 670} (2003) 439, {\tt hep-th/0307042}.

\bibitem{spinning} N.~Beisert, J.~A.~Minahan, M.~Staudacher and K.~Zarembo,
{\it Stringing spins and spinning strings},
JHEP {\bf 0309} (2003) 010, {\tt hep-th/0306139}.

\bibitem{spectroscopy} N.~Beisert, S.~Frolov, M.~Staudacher and A.~A.~Tseytlin,
{\it Precision spectroscopy of AdS/CFT},
JHEP {\bf 0310} (2003) 037, {\tt hep-th/0308117}.

\bibitem{AS03} G.~Arutyunov and M.~Staudacher,
{\it Matching higher conserved charges for strings and spins},
JHEP {\bf 0403} (2004) 004, {\tt hep-th/0310182}.

\bibitem{duals} J.~Engquist, J.~A.~Minahan and K.~Zarembo,
{\it Yang-Mills duals for semiclassical strings on $AdS_5 \times S^5$},
JHEP {\bf 0311} (2003) 063, {\tt hep-th/0310188}.

\bibitem{SS} D.~Serban and M.~Staudacher,
{\em Planar ${\cal N} = 4$ gauge theory and the Inozemtsev long range spin chain},
JHEP {\bf 0406} (2004) 001, {\tt hep-th/0401057}.

\bibitem{Kristjansen} C.~Kristjansen,
{\em Three-spin strings on $AdS_5 \times S^5$ from ${\cal N} = 4$ SYM}, \hfill\break
Phys.\ Lett.\ B {\bf 586} (2004) 106, {\tt hep-th/0402033}.

\bibitem{Engquist} J.~Engquist,
{\em Higher conserved charges and integrability for spinning strings in $AdS_5 \times S^5$},
JHEP {\bf 0404} (2004) 002, {\tt hep-th/0402092}.

\bibitem{Arutyunov} G.~Arutyunov and M.~Staudacher,
{\em Two-loop commuting charges and the string / gauge duality}, {\tt hep-th/0403077}.

\bibitem{Dimov1} H.~Dimov and R.~C.~Rashkov,
{\em A note on spin chain / string duality},
{\tt hep-th/0403121}.

H.~Dimov and R.~C.~Rashkov,
{\em Generalized pulsating strings}, JHEP {\bf 0405} (2004) 068,
{\tt hep-th/0404012}.

\bibitem{Ferretti} G.~Ferretti, R.~Heise and K.~Zarembo,
{\em New integrable structures in large-N QCD}, Phys.\ Rev.\ D {\bf 70} (2004) 074024, 
{\tt hep-th/0404187}.

\bibitem{BDS} N.~Beisert, V.~Dippel and M.~Staudacher,
{\em A novel long range spin chain and planar ${\cal N} = 4$ super Yang-Mills},
JHEP {\bf 0407} (2004) 075, {\tt hep-th/0405001}.

\bibitem{Smedback} M.~Smedb\"ack,
{\em Pulsating strings on $AdS_5 \times S^5$},
JHEP {\bf 0407} (2004) 004, \hfill\break {\tt hep-th/0405102}.

\bibitem{Freyhult} L.~Freyhult,
{\em Bethe ansatz and fluctuations in $SU(3)$ Yang-Mills operators},
JHEP {\bf 0406} (2004) 010, {\tt hep-th/0405167}.

\bibitem{beyondsu2} J.~A.~Minahan,
{\em Higher loops beyond the $SU(2)$ sector}, 
JHEP {\bf 0410}, 053 (2004), {\tt hep-th/0405243}.

\bibitem{Kristjansensu3} C.~Kristjansen and T.~Mansson,
{\em The circular, elliptic three-spin string from the SU(3) spin chain},
Phys.\ Lett.\ B {\bf 596} (2004) 265, {\tt hep-th/0406176}.

\bibitem{AFS} G.~Arutyunov, S.~Frolov and M.~Staudacher,
{\em Bethe ansatz for quantum strings}, JHEP {\bf 0410} (2004) 016, {\tt hep-th/0406256}.

\bibitem{Ryang} S.~Ryang,
{\em Circular and folded multi-spin strings in spin chain sigma models}, 
JHEP {\bf 0410} (2004) 059, {\tt hep-th/0409217}.

\bibitem{Kazakov} V.~A.~Kazakov, A.~Marshakov, J.~A.~Minahan and K.~Zarembo,
{\em Classical / quantum integrability in AdS/CFT},
JHEP {\bf 0405} (2004) 024, {\tt hep-th/0402207}.

\bibitem{SL2curve} V.~A.~Kazakov and K.~Zarembo,
{\em Classical / quantum integrability in non-compact sector of AdS/CFT},
JHEP {\bf 0410} (2004) 060, {\tt hep-th/0410105}.

\bibitem{Beisert:2004ag} N.~Beisert, V.~A.~Kazakov and K.~Sakai,
{\em Algebraic curve for the SO(6) sector of AdS/CFT},
{\tt hep-th/0410253}.

\bibitem{ArutyunovF} G.~Arutyunov and S.~Frolov,
{\em Integrable Hamiltonian for classical strings on $AdS_5 \times S^5$},
 JHEP {\bf 0502} (2005) 059, {\tt hep-th/0411089}.

\bibitem{n=4curve} S.~Schafer-Nameki,
{\em The algebraic curve of 1-loop planar ${\cal N} = 4$ SYM}, \hfill\break
{\tt hep-th/0412254}.

\bibitem{Martin} M.~Kruczenski, {\it Spin chains and string theory},
Phys.\ Rev.\ Lett.\  {\bf 93} (2004) 161602, {\tt hep-th/0311203}.

\bibitem{KRT} M.~Kruczenski, A.~V.~Ryzhov and A.~A.~Tseytlin,
{\em Large spin limit of $AdS_5 \times S^5$ string theory and low energy expansion of
ferromagnetic spin chains}, Nucl.\ Phys.\ B {\bf 692} (2004) 3, {\tt hep-th/0403120}.

\bibitem{su3} R.~Hern\'andez and E.~L\'opez,
{\em The $SU(3)$ spin chain sigma model and string theory},
JHEP {\bf 0404} (2004) 052, {\tt hep-th/0403139}.

\bibitem{Stefanski} B.~J.~Stefanski and A.~A.~Tseytlin,
{\em Large spin limits of AdS/CFT and generalized Landau-Lifshitz equations},
JHEP {\bf 0405} (2004) 042, {\tt hep-th/0404133}.

\bibitem{KT} M.~Kruczenski and A.~A.~Tseytlin,
{\em Semiclassical relativistic strings in $S^5$ and long coherent operators in 
${\cal N}=4$ SYM theory}, JHEP {\bf 0409} (2004) 038, {\tt hep-th/0406189}.

\bibitem{sl2} S.~Bellucci, P.~Y.~Casteill, J.~F.~Morales and C.~Sochichiu,
{\em $SL(2)$ spin chain and spinning strings on $AdS_5 \times S^5$},
Nucl.\ Phys.\ B {\bf 707} (2005) 303, 
{\tt hep-th/0409086}.

\bibitem{fermions} R.~Hern\'andez and E.~L\'opez,
{\em Spin chain sigma models with fermions},
JHEP {\bf 0411} (2004) 079, {\tt hep-th/0410022}.

\bibitem{LZ} M.~Lubcke and K.~Zarembo,
{\em Finite-size corrections to anomalous dimensions in N = 4 SYM theory},
JHEP {\bf 0405} (2004) 049, {\tt hep-th/0405055}.

\bibitem{semi} S.~Frolov and A.~A.~Tseytlin,
{\em Quantizing three-spin string solution in $AdS_5 \times S^5$}, 
JHEP {\bf 0307} (2003) 016, {\tt hep-th/0306130}.

\bibitem{FPT} S.~A.~Frolov, I.~Y.~Park and A.~A.~Tseytlin,
{\em On one-loop correction to energy of spinning strings in $S^5$}, 
Phys.\ Rev.\ D {\bf 71} (2005) 026006, {\tt hep-th/0408187}.

\bibitem{SL2} I.~Y.~Park, A.~Tirziu and A.~A.~Tseytlin,
{\em Spinning strings in $AdS_5 \times S^5$: One-loop correction to energy in $SL(2)$
sector}, {\tt hep-th/0501203}.

\bibitem{FK} L.~Freyhult and C.~Kristjansen,
{\em Finite Size Corrections to Three-spin String Duals},
{\tt hep-th/0502122}.

\bibitem{R-limit} R. W. Richardson, {\em Pairing in the limit
of a large number of particles}, J. Math. Phys. {\bf 18} (1977) 1802. 

\bibitem{G-book} M. Gaudin, ``\'Etats propres et valeurs propres
de l'Hamiltonien d'appariement'', unpublished Saclay preprint, 1968.
Included in  Travaux de Michel Gaudin, Mod\`eles exactament r\'esolus, 
Les \'Editionsde Physique, France, 1995.

\bibitem{largeN} J.~M.~Rom\'an, G.~Sierra and J.~Dukelsky,
{\em Large N limit of the exactly solvable BCS model: analytics versus
numerics}, Nucl.\ Phys.\ B {\bf 634} (2002) 483, {\tt cond-mat/0202070}.

\bibitem{RMP} J.~Dukelsky, S.~Pittel and G.~Sierra,
{\em Exactly solvable Richardson-Gaudin models for many-body quantum systems},
Rev. Mod. Phys. {\bf 76} (2004) 643-662, 
{\tt nucl-th/0405011}.

\bibitem{RS} R.W. Richardson and N. Sherman,
{\em Exact eigenstates of the pairing-force Hamiltonian},
Nucl. Phys. {\bf B 52} (1964) 221.

\bibitem{R-boson} R.W. Richardson, 
{\em Exactly Solvable Many-Boson Model},
J. Math. Phys. {\bf 9} 1327 (1968). 

\bibitem{Jorge-boson} J.~Dukelsky and P.~Schuck,
{\em Condensate Fragmentation in a New Exactly Solvable Model for Confined
Bosons}, Phys.\ Rev.\ Lett.\  {\bf 86} (2001) 4207, {\tt cond-mat/0009057}.

\bibitem{RD1} A.~LeClair, J.~M.~Rom\'an and G.~Sierra,
{\em Russian Doll Renormalization Group and Superconductivity},
Phys.\ Rev.\ B {\bf 69} (2004) 20505, {\tt cond-mat/0211338}.

\bibitem{RD2} A. Anfossi, A. LeClair and G. Sierra, {\em The elementary excitations 
of the exactly solvable russian doll BCS model of superconductivity},
to appear. 

\bibitem{Tseytlin:2003ii} A.~A.~Tseytlin,
{\em Spinning strings and AdS/CFT duality},
{\tt hep-th/0311139}.

\bibitem{bosonBAE} A.~A.~Ovchinnikov,
{\em On the exactly solvable pairing models for bosons},
JSTAT {\bf 0407} (2004) P004, {\tt math-ph/0405047}.

\bibitem{cardy} J. Cardy, ``Scaling and Renormalization
in Statistical Physics'', Cambridge University Press, U.K. (1997). 

\bibitem{boson} A. O. Gogolin, A.A. Nersesyan and A.M. Tsvelik,
``Bosonization and Strongly Correlated Systems'', 
Cambridge University Press, U.K. (1998).

\bibitem{BTZ} N.~Beisert, A.~A.~Tseytlin and K.~Zarembo, {\em Matching quantum strings 
to quantum spins: one-loop vs. finite size corrections}, 
{\tt ITEP-TH-12/05, PUTP-2151, UUITP-02/05, hep-th/0502173}.


\end{thebibliography}
\end{document}